\documentclass[sigconf]{acmart}
\usepackage[colorinlistoftodos]{todonotes}
\usepackage{verbatim}
\usepackage{tabularx}
\usepackage{soul}
\usepackage{enumitem}
\usepackage{booktabs}
\usepackage{blindtext}
\usepackage[labelformat = empty]{subfig}
\usepackage[font=small,skip=3pt]{caption}

\def\BibTeX{{\rm B\kern-.05em{\sc i\kern-.025em b}\kern-.08emT\kern-.1667em\lower.7ex\hbox{E}\kern-.125emX}}
    
\copyrightyear{2019}
\acmYear{2019}
\setcopyright{acmlicensed}

\graphicspath{ {figures/} }
\begin{document}

%
\title{Product Insights: Analyzing Product Intents in Web Search}

%
\author{Nikitha Rao}
\email{t-nirao@microsoft.com}
\affiliation{%
  \institution{Microsoft Research}
  \streetaddress{}
  \city{Bangalore}
  \state{India}
}

\author{Chetan Bansal}
\email{chetanb@microsoft.com}
\affiliation{%
  \institution{Microsoft Research}
  \streetaddress{}
  \city{Redmond}
  \state{WA, USA}
}

\author{Subhabrata Mukherjee}
\email{submukhe@microsoft.com}
\affiliation{%
  \institution{Microsoft Research}
  \streetaddress{}
  \city{Redmond}
  \state{WA, USA}
}

\author{Chandra Maddila}
\email{chmaddil@microsoft.com}
\affiliation{%
  \institution{Microsoft Research}
  \streetaddress{}
  \city{Redmond}
  \state{WA, USA}
}

%

%
\begin{abstract}
Web search engines are frequently used to access information about products. This has increased in recent times with the rising popularity of e-commerce. However, there is limited understanding of what users search for and their intents when it comes to product search on the web. In this work, we study search logs from Bing web search engine to characterize user intents and study user behavior for product search. We propose a taxonomy of product intents by analyzing product search queries. This is a challenging task given that only $15\%-17$\% of web search queries are about products. We train machine learning classifiers with query log features to classify queries based on intent with an overall $F_1$-score of $78\%$. We further analyze various characteristics of product search queries in terms of search metrics like dwell time, success, popularity and session-specific information.
\end{abstract}

%
%


%
\maketitle
\section{Introduction}

Product search refers to the user intent of accessing product-related information from the web. A number of these searches are geared for online shopping or accessing e-commerce websites. As per a recent study~\cite{Manufacturing.Net}, $7$ out of $10$ individuals in the U.S. purchased at least $1$ product online in $2018$ with a total spending of $\$500$ billion. 
According to various surveys\footnote{https://blog.survata.com/amazon-takes-49-percent-of-consumers-first-product-search-but-search-engines-rebound}, Amazon.com accounts for half of the product search queries in the web. However, traditional search engines are still popular among internet users when they do not have a specific purchase in mind. For instance, nearly $46\%$ of such users will start on search engines compared to $39\%$ for Amazon. Therefore, it is useful to characterize and better understand user intent for product search on the web. 

Query understanding for general web search and user intent has been extensively studied in the past for domains like healthcare~\cite{inproceedings}, developer behavior \cite{bansal2019usage}, employment~\cite{chancellor2018measuring} and security ~\cite{bansal2020studying}. 
One of the earliest works in this space is by Broder~\cite{article} who proposed a taxonomy to categorize web search into Navigational, Informational and Transactional queries. This work was followed up by Rose and Levinson~\cite{Rose04understandinguser} who created a framework for understanding user goals. In the e-commerce space, Moe et al.~\cite{articleMoe} perform a study to better understand online shoppers' behavior by classifying their intents into categories like buying, browsing, searching and knowledge-building, etc. This work is followed by Su et al.~\cite{su2018user} who worked on building a taxonomy of product-related search queries by conducting a controlled user study within an e-commerce website. There has also been some work done on improving product search within e-commerce websites~\cite{long2012enhancing, guo2018multi} primarily with regards to the purchase intent. In contrast, our work focuses on general web search queries with a broader set of applicable user intents.

In this work, we want to answer the following questions:
\begin{itemize}
    \item[RQ1.] What percentage of web search queries are related to product search?
    \item[RQ2.] What are the different user intents for product search? 
    \item[RQ3.] How does product search intent correlate with search metrics like popularity, success and dwell time, and what are their underlying characteristics? 
\end{itemize}

A study~\cite{li2008learning} performed a decade back shows that approximately 5\%-7\% of the distinct web search queries contain product intent. With RQ1 (discussed in Section~\ref{sec:product-dataset}), we want to validate if that still holds in recent times given the change induced by e-commerce search engines. Traditional web search queries have intents like {\em Navigational}, {\em Informational} and {\em Transactional}~\cite{article}. With RQ2 (discussed in Section~\ref{sec:taxonomy}), we investigate if product search has similar or additional intents. 
With RQ3 (in Section~\ref{sec:analysis}), we want to study the characteristics and distribution of different product search intents. 

Consequently, we analyze user search behavior from a random sample of 1 million search queries collected from the first week of September 2019 from a popular web search engine. A large number of user queries are found to be not related to products upon inspection. Therefore, we develop a machine learning model to identify product-related queries. We then build a product search intent taxonomy following open coding technique and human annotation as used in prior works~\cite{jansen} for web search intent categorization. Finally, we develop an intent classification model to further classify the product-related queries to one of several intent types leveraging various query log attributes as features. 
We apply our models to large-scale web search query logs to find the distribution of different intents and analyze how they correlate with several search metrics. We discover several interesting insights like Transactional queries are the most popular whereas Navigational queries have the highest rate of success (in Section \ref{sec:analysis}). 

\section{Product Query Classification}
\label{sec:product-classifier}
In this paper, we want to analyze and characterize how web search is used for product search. In order to do so, we first need to distinguish product search queries from general search queries. One possibility is to manually label a large number of queries to find names of different products and variations. This is unlikely to give a lot of true positives given that only 5\%-7\% of the distinct web search queries are likely to be product queries from prior research~\cite{li2008learning}. 
To alleviate this problem, we leverage various search engine features to identify product-related queries that can be used to train machine learning models to identify similar related queries.

\noindent \textbf{Distant supervision heuristics}: We explore and evaluate the following heuristics to automatically collect a large amount of positive examples for product-related queries to train the classifiers. Negative examples for non-product queries are randomly sampled from the remaining query set.

\begin{enumerate}[leftmargin=3mm,itemsep=0mm]
    \item \textbf{Product Ads}: Most of the major search engines such as Google and Bing allow advertisers to create Ad campaigns \cite{GoogleShopping} for queries related to products. We label those queries as product-specific for which one or more product Ads were displayed. Note that this technique implicitly uses the Ad recommendation and ranking algorithm of the search engine.
    
    \item \textbf{Product Categories}: We consider another heuristic that relies on the list of product categories from Amazon.com \cite{AmazonList}. These categories (such as, automotive, books, etc.) cover the vast majority of products available therein. We label each query as product related if the search query or the click url contains any of the product category names.
    
    \item \textbf{Product Ads and Categories}: This uses a combination of the previous two heuristics.
    
    \item \textbf{Product list}: Finally, we also use the top 10 best selling products of all time \footnote{https://time.com/92765/10-best-selling-products-ever/}. We classify any query where the query text or the click url contains any of these products as a positive example. 
\end{enumerate}

\noindent \textbf{Dataset}: We want to select the best heuristic from the above list to generate our training data. Subsequently, we randomly sampled $500$ search queries from the query logs and manually labeled them as {\em product} or {\em non-product} queries.  We found that 17.4\% of these queries were product queries. 
Table \ref{product-heuristic} presents the evaluation of all the heuristics on our manually annotated query set. We observe that the \textit{Product Ads and Categories} heuristic is the most accurate one with $71\%$ $F_1$-score for the product class and $91\%$ accuracy overall. Now, we generate the training data by sampling $500k$ queries that satisfy this heuristic as positive examples and another $500k$ queries that do not as negative examples.

\noindent \textbf{Product query classifier}:  
The search query and click urls from the training dataset are tokenized and pre-trained Word2Vec \cite{Mikolov:2013:DRW:2999792.2999959} embeddings are used to generate distributed representations of these features. Both the query and the click urls are represented in a $300$ dimensional embedding space. Word embeddings have been shown to improve performance in many classification tasks \cite{lilleberg2015support, tang2014learning}. We train several classifiers and report five fold cross validation results in Table \ref{product-manual}. Among all the classifiers, Multi-Layer Perceptron obtain the best performance with an $F_1$-score of $84\%$ and used for the remaining analysis. 


\begin{table}
\small

\caption{Heuristics evaluation for distant supervision.}
\label{product-heuristic}
\setlength\belowcaptionskip{7pt}
 \begin{tabular}{lcccc} 
 \toprule
 \textbf{Heuristic} & \textbf{Precision} & \textbf{Recall} & \textbf{F1-score} & \textbf{Accuracy} \\
 \midrule
 Product List & 57 & 5 & 9 & 83 \\
 Product Ads & 74 & 54 & 62 & 89 \\  
 Product Categories & \textbf{83} & 24 & 37 & 86 \\  
 \textbf{Ads \& Categories} & 74 & \textbf{68} & \textbf{71} & \textbf{91} \\\bottomrule
\end{tabular}
\end{table}

\begin{table}
\small
\caption{ Five-fold cross validation of product query classifiers.}
\label{product-manual}

\setlength\belowcaptionskip{7pt}
 \begin{tabular}{lcccc} 
\toprule
 \textbf{Classifier} & \textbf{Precision} & \textbf{Recall} & \textbf{F1-score} & \textbf{Accuracy} \\
 \midrule
 AdaBoost & 76 & 74 & 75 & 75 \\
 LinearSVC & 82 & 77 & 79 & 79 \\  
 LogisticRegression & 82 & 77 & 79 & 79 \\  
 \textbf{MLP} & \textbf{83} & \textbf{84} & \textbf{84} & \textbf{83} \\  
 \bottomrule
\end{tabular}
\end{table}

\section{Product Query Intent Taxonomy}
\label{sec:taxonomy}
Users can have different intents behind product-related searches. They might be looking for some information about a product, compare them to other similar items, make a purchase and so on. In this section, we study the various user intents for product search.


\subsection{Dataset generation}
\label{sec:product-dataset}
We use the best performing product query classification model (multi-layer perceptron) on a random sample of 1 million search queries, collected from the first week of September 2019, and obtain $149,579$ product-related queries which implies that approximately $15\%$ of all search queries in the web are product searches. We found this number to be around $17\%$ on the manually labeled set of $500$ randomly sampled queries in Section~\ref{sec:product-classifier}. Our findings demonstrate that product search queries account for $15\%-17\%$ which is significantly higher than the study~\cite{li2008learning} from a decade ago which found the volume to be $5\%-7\%$. 
For the next phase of analysis, we use the above classified product-related queries along with other query features from the search engine like the click urls, click snippets, click counts, etc. that provide more context about the search intent.

\begin{table*}[t!]
\small
  \begin{center}
    \caption{Intent categories for product search queries with distribution as per manual annotation on $1500$ samples.}
    
    \label{tab:taxonomy}
    \begin{tabular}{p{2.5cm}p{7cm}p{6cm}}

     \toprule
      \textbf{Intent \newline (Distribution)} & \textbf{Description} & \textbf{Examples} \\
   \midrule
      Not Product Related \newline (42.50\%) & Queries which are not product related. & `apple stock price', `homedepot jobs', `walmart customer relations address' \\
   \midrule      
      Comparison \newline (11.75\%) & Queries for comparing products in a category or based on attributes such as features, price, etc. &  `skype vs microsoft teams', `ironman action figures', `compare iphone x and xr'  \\
   \midrule      
      Informational \newline (15.56\%) & User is looking for some specific product-related information. & `what does apple care cover on watches', `sprinkler system design', `where to buy battens?' \\
   \midrule      
      Navigational \newline (11.19\%) & User wants to navigate to a specific website associated with a product. & `itunes sign in', `amazon chime', `apple care', `myamazon/kdp', `alexa apps' \\
   \midrule      
      Support \newline (8.88\%) & Here the user is looking for help to troubleshoot some product specific issue. & `hp mouse not working', `cannot connect to display mininet', `access denied amazon aws' \\
   \midrule      
      Transactional \newline (10.12\%) & Such queries depict that the user is interested in transactions like buying, downloading or installing a specific product. & `antique german wind up car', `qvc gold jewelry clearance diamonique', `h.h. scott S 10 speakers for sale'\\
   \bottomrule      
    \end{tabular}
  \end{center}
\end{table*}

\subsection{Clustering and sampling for annotation} 
Given a large number of queries from the previous step, we want to sample a set of queries for manual annotation. Random sampling of queries result in the risk of missing those from less popular or scarcely appearing intents. To explore queries from diverse intents, we 
leverage Latent Dirichlet Allocation (LDA)~\cite{blei2003latent} to cluster textually similar queries and sample queries from each cluster. 

LDA assumes a document (query in our case) to have a distribution over latent topics (intent in our case), and each topic to have a distribution over words. We use wordpiece tokenization~\cite{DBLP:journals/corr/WuSCLNMKCGMKSJL16} to tokenize the query string, clicked urls and click snippets to get all the keywords associated with the query that are concatenated to form the input document for the LDA model. The clicked urls and snippets provide additional context to the query. The tokenization considers a fixed vocabulary size (e.g., $30k$ tokens) and segments words into wordpieces such that every piece is present in the vocabulary. A special symbol `\#\#' is used to mark the split of a word into its sub-pieces. We remove the domain information from the feature set to prevent trivial clustering based on domain name. 
The length of the feature vector is $30k$ given by wordpiece vocabulary size. We set the number of latent topics to $50$. We then randomly sample $30$ queries from each topic cluster resulting in $1500$ samples that are used for manual annotation.

\subsection{Manual annotation}
Three annotators manually inspected all the queries collected in the previous step along with all the relevant contextual features. Each annotator was asked the following questions for each query:
\begin{enumerate}
    \item Is the query related to a product?
    \item For product-related queries, does any of the following intent categories apply, {\em Informational}, {\em Navigational}, {\em Transactional}?
    \item Is there any other intent for product-related queries that is not included in the above intent categories?
\end{enumerate}

The intents corresponding to {Informational}, {Navigational} and {Transactional} are commonly observed in web search queries~\cite{article}.  In addition to those, the annotators observed (i) a large number of product queries for comparing various products based on attributes such as features and price, and (ii) a large number of support-related queries where the user is looking for help to troubleshoot some product specific issue. Given these observations, the annotators were asked to re-annotate such queries with the intent categories corresponding to {\em Comparison} and {\em Support}. We ignored queries in languages other than English. This resulted in five intent categories for product-related queries. Table~\ref{tab:taxonomy} shows a detailed description for each intent category with examples and key indicators from our analysis. Introducing product specific intents like Comparison and Support can help improve the quality of search results for product queries. 
We computed the inter-annotator agreement using Fleiss kappa ~\cite{fleiss1971measuring} as $0.79$ indicating substantial agreement among the annotators for different intent categories. Note that the distribution of intents in Table~\ref{tab:taxonomy} is computed over the manually annotated set of $1500$ queries and does not reflect the true distribution of intents in the wild. An analysis of the true distribution of intents has been carried out in Section~\ref{sec:analysis}.

In Table~\ref{tab:prior_taxonomy}, we show a comparison of user search intent taxonomies as developed in prior works including general web search~\cite{article,Rose04understandinguser}, product search within e-commerce search engines~\cite{su2018user}, decision making by consumers for buying goods and services~\cite{BuyerDecisionProcess} and search in the context of interactive applications~\cite{Fourneyinproceedings}. We observe significant overlap in the taxonomy of all these search intents. Our newly introduced product search intent categories, namely comparison and support overlap with the {\em evaluation of alternatives} and {\em post purchase behaviour} from the buyer decision process~\cite{BuyerDecisionProcess}.

\begin{table}
\small

  \begin{center}
    \caption{Taxonomy of user search intent in prior work.}
    \label{tab:prior_taxonomy}
    \begin{tabular}{p{2cm}p{5.9cm}}
     \toprule
        \textbf{Reference} & \textbf{Taxonomy} \\
     \midrule
        Broder et al.~\cite{article} & Informational, Navigational, Transactional\\
       Buyer Decision Process~\cite{BuyerDecisionProcess} & Problem/Need Recognition, Information Search, Evaluation of Alternatives, Purchase Decision, Post Purchase Behavior \\
       Su et al.~\cite{su2018user} & Target Finding, Decision Making, Exploration \\
       Rose et al.~\cite{Rose04understandinguser} & Navigational, Informational (Directed, Undirected, Advice, Locate, List), Resource (Download, Entertainment, Interact, Obtain) \\
       Fourney et al.~\cite{Fourneyinproceedings} & Operation Instruction, Troubleshooting, Reference, Download, General Information, Off-Topic  \\
        This work & Informational, Navigational, Transactional, Comparison, Support, Not-product-related \\
   \bottomrule      
    \end{tabular}
  \end{center}
\end{table}

\begin{figure*}[h]
\caption{Feature distribution across different intents.}
\begin{center}
\subfloat[][]{\includegraphics[width=0.33\textwidth]{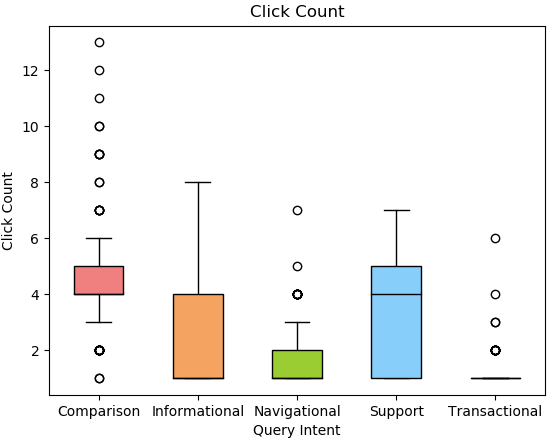}
\includegraphics[width=0.33\textwidth]{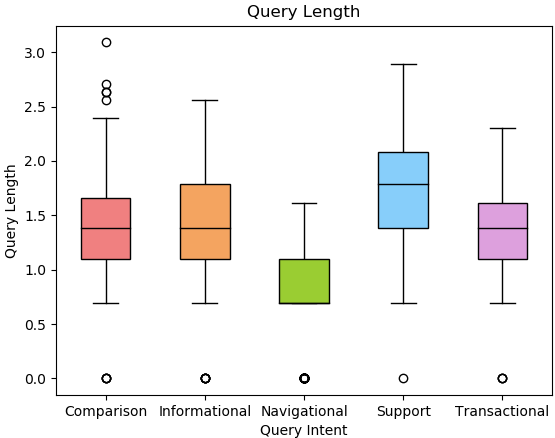}
\includegraphics[width=0.33\textwidth]{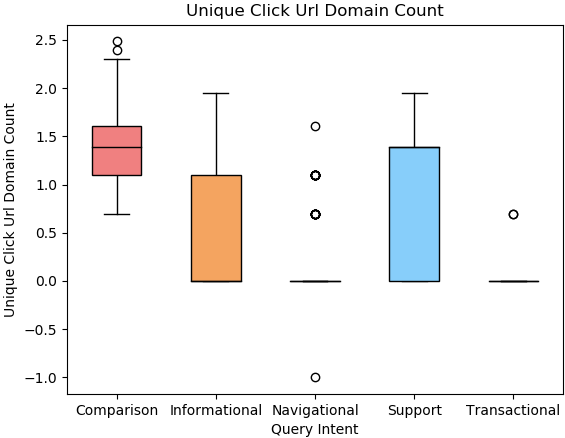}}
\vspace{-4mm}
\subfloat[]{\includegraphics[width=0.33\textwidth]{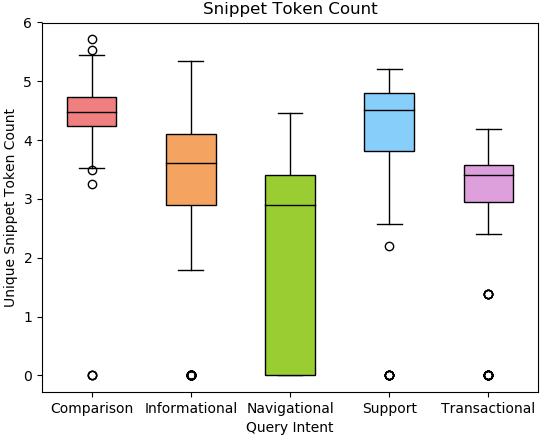}
\includegraphics[width=0.33\textwidth]{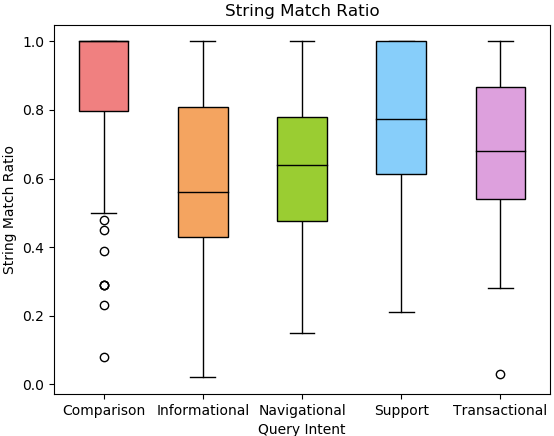}}
\end{center}
\label{fig:features}
\vspace{-4mm}
\end{figure*}

\subsection{Intent Classification}
\label{sec:classifier}

Given the set of $1500$ manually annotated queries with product search intents, we aim to train machine learning models to automatically perform intent classification of user queries.

\vspace{3mm} 
\noindent{\bf Features and preliminary analysis}: We use the following query log features for intent classification. Figure \ref{fig:features} shows a distribution of these features over different intent categories in our labeled data. 
\vspace{-2mm}
\begin{itemize}

\item[a.]{\em Query embeddings}: Average of the individual (wordpiece tokenized) word embeddings from pre-trained Word2Vec~\cite{Mikolov:2013:DRW:2999792.2999959} model.

\item[b.]{\em Url embeddings}: Similar to query embeddings, we tokenize and aggregate the word embeddings for the clicked urls.

\item[c.]{\em Click Count}: Number of urls the user has clicked on. From Figure~\ref{fig:features}, we observe that Transactional queries mostly have a single click depicting a crisp user objective. Navigational queries have one or two clicks where the user can easily recognize the website they want to navigate to from the suggestions. On the other hand, Support, Comparison and Informational queries tend to have a much higher click count since these intents are more exploratory in nature. 

\item[d.]{\em Query length}: Number of tokens present in the query string. From Figure~\ref{fig:features}, we observe that Navigational queries are significantly shorter in length; whereas queries with Support intent are significantly longer as they tend to be more descriptive.

\item[e.]{\em Snippet token count}: Number of tokens present in the click snippets. From Figure~\ref{fig:features}, we observe that the number of snippet tokens is much higher for Support, Comparison and Informational queries, since the clicked websites have much higher textual content when compared to Navigational and Transactional queries.    

\item[f.]{\em Url domain count}: Number of unique domains in the clicked urls. From Figure~\ref{fig:features}, we observe that the count is at most two for Navigational queries as the user knows which website they want to navigate to. The url domain count is much higher for Comparison queries as the user compares the product feature or price across multiple websites. Similarly, Informational and Support queries are more exploratory in nature.

\item[g.]{\em Similarity}: Ratio of unique tokens that are present in the query string to those in the clicked urls. This ratio is higher for Support queries where the query and clicked urls are more descriptive of the issue the user is facing. Similarly, for Comparison queries most of the query tokens are contained in the clicked urls. Navigational and Transactional queries tend to be much shorter resulting in a low ratio. Similarly, Informational queries are more exploratory in nature and have a low ratio.
\end{itemize}

\noindent{\bf Model}: For each intent, we use $80\%$ of the labeled data for training, $20\%$ for test and report five fold cross validation results. We train different classification models and evaluate them based on accuracy, precision, recall and $F_1$ scores, as reported in Table~\ref{tab:intentscores}. We observe that Linear Support Vector Machines perform the best across all the classes with an overall accuracy of $77\%$ and $F_1$ score of $78\%$.



\begin{table*}[t!]
\small
  \begin{center}

    \caption{Comparison of various models for intent classification with five-fold cross validation. }
    \label{tab:intentscores}
    \begin{tabular}{c|c|ccc|ccc|ccc|ccc|ccc}
    \toprule
      \textbf{Classifier}  & \textbf{Test} & \multicolumn{3}{|c|}{\textbf{Comparison}} & \multicolumn{3}{|c|}{\textbf{Informational}} & \multicolumn{3}{|c|}{\textbf{Navigational}} & \multicolumn{3}{|c|}{\textbf{Support}} & \multicolumn{3}{|c}{\textbf{Transactional}} \\\cline{3-17}
       &  \textbf{Accuracy} & \textbf{P } & \textbf{R } & \textbf{ F1 } &\textbf{P } & \textbf{R } & \textbf{ F1} & \textbf{P} & \textbf{R} & \textbf{F1} & \textbf{P} & \textbf{R} & \textbf{F1} & \textbf{P} & \textbf{R} & \textbf{ F1}\\\midrule
    Random Forest & 64.67 & 75 & 88 & 81 & 48 & 49 & 48 & 74 & 71 & 73 & 67 & 64 & 65 & 65 & 53 & 58 \\
    \textbf{Linear SVM} & \textbf{77.17} &  \textbf{91} & 84 & \textbf{88} & 69 & \textbf{70} & \textbf{70} & \textbf{78} & 81 & \textbf{79} & 75 & 75 & 75 & 76 & \textbf{78} & 77 \\
    Gaussian SVM & 70.65 & 66 & \textbf{95} & 78 & \textbf{72} & 42 & 53 & 67 & 81 & 73 & \textbf{79} & \textbf{86} & \textbf{83} & 75 & 77 & 76 \\
    K-Nearest Neighbors & 74.45 & 83 & 70 & 76 & 57 & 53 & 55 & 62 & \textbf{86} & 72 & 51 & 71 & 60 & \textbf{96} & 66 & \textbf{78} \\
    \bottomrule
    \end{tabular}
  \end{center}
\end{table*}

\vspace{-2mm}
\section{Query Intent Analysis}
\label{sec:analysis}

We apply the best intent classification model (linear SVM) from the previous section to analyze the product queries (described in Section~\ref{sec:product-dataset}) based on several measures like popularity, success rate and effort estimation. We also do a session-wise analysis to gain insights on what intents co-occur within a given session.

\noindent {\bf Intent success rate}: A search is said to be successful if the user spends more than 30 seconds on the clicked url page as per a study in~\cite{fox2005evaluating}. Since a user may click on multiple urls during product search, we consider a search to be successful if the dwell time on the last clicked url is more than 30 seconds. 
From Table \ref{tab:popularity}, we observe that Navigational queries have the highest success rate of 77.28\% whereas Comparison queries are least successful.

\noindent{\bf Intent popularity} is calculated as the percentage of product queries having a specific intent out of all query intents. From Table \ref{tab:popularity} we observe that the most popular intent is Transactional followed by Navigational; whereas Comparison and Support queries are less popular ones.

\noindent {\bf Intent Effort Estimation}: The effort taken to complete a search is proportional to the dwell time. The values are then transformed to be relative to the comparison intent. From Table~\ref{tab:popularity}, we observe that the effort put into Comparison, Support and Transactional query is much less compared to that for Informational and Navigational query. This results from the former intents (e.g., queries like `skype vs microsoft teams' and `iphone x price') being more task-specific compared to the latter exploratory intents (e.g., queries like `sprinkler system design' and `alexa apps').

\begin{table}
\small
\caption{Comparison of product intent metrics.}
    \label{tab:popularity}
\begin{tabular}{cccc}
     \toprule
     \textbf{Intent} & \textbf{Success Rate}  & \textbf{Popularity} & \textbf{Estimated Effort}\\
     \textbf{} & \textbf{(\%)}  & 
     \textbf{(\%)} & \textbf{(Relative Scale)}\\
    \midrule
     Comparison & 61.31 & 0.27 & 1\\ 
     Informational & 70.97 & 21.72 & 1.7 \\ 
     Navigational & 77.28 & 25.71 & 2.63 \\ 
     Support & 68.95 & 1.28 & 1.44 \\ 
     Transactional & 64.69 & 51.03 & 1.43 \\ 
    \bottomrule
    \end{tabular}
\end{table}

\begin{figure}
    \caption{Intent Co-occurrence Distribution.}
    \includegraphics[width=0.5\textwidth]{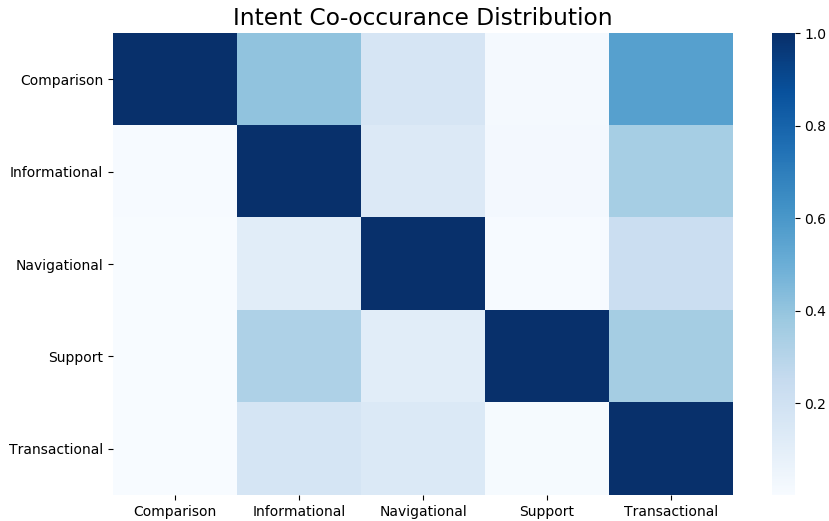}
    \label{fig:cooccurrence}
\end{figure}

\noindent {\bf Intent Co-occurrence}: We analyzed the temporal co-occurrence of different intents in a given session aggregated across all the sessions. From Figure~\ref{fig:cooccurrence} we observe that while less than $1$\% of Comparison and Support queries are preceded by other intents, $56$\% of the Transactional queries are preceded by Comparison queries indicating that many users indeed compare products before making a purchase. Comparison is also often followed by Informational  ($41$\%) queries depicting that the user is looking for more information about a product after shortlisting the choice. Another interesting insight we observed is that Support queries are commonly followed by Informational ($33$\%) and Transactional ($36$\%) queries indicating that users seek additional information to help understand the solution (i.e. troubleshoot) or purchase a replacement for the product.

\section{Conclusion}
\label{sec:conclusion}
In this work, we performed an extensive study to characterize product search queries and their intents from query log analysis of Bing web search engine. This is in contrast to prior work that heavily focused on e-commerce search engines with limited diversity of product search intents. 
We developed the product query intent taxonomy for web search and trained machine learning models with a judicious selection of features from search engine and query context to automatically identify product search intents. We applied our intent classification models to large-scale query logs and reported our findings from the intent distribution analysis.


%
\bibliographystyle{ACM-Reference-Format}
\bibliography{references}

\end{document}